\documentclass{elsart}%
\usepackage{amsfonts}
\usepackage{amsmath}
\usepackage{amssymb}
\usepackage{graphicx}%
\begin{document}
\begin{frontmatter}
\title{Scaling and universality in the micro-structure of urban space}
\author{Rui Carvalho \thanksref{label1}}
\thanks[label1]{Corresponding author. Email address: rui.carvalho@ucl.ac.uk}
\collab{Alan Penn}
\address{The Bartlett  School of Graduate Studies\\
1-19 Torrington Place\\
University College London, Gower Street\\
London WC1E 6BT, United Kingdom}
\begin{abstract}
We present a broad, phenomenological picture of the distribution of the length of open space linear
segments, $l$, derived from maps of $36$ cities in $14$ different countries.
By scaling the Zipf plot of $l$, we obtain two master curves for a sample of cities, which are not a function of city size. We show that a
third class of cities is not easily classifiable into these two
universality classes. The cumulative distribution of $l$ displays power-law tails with two distinct exponents, $\alpha_B=2$ and $\alpha_R=3$.
We suggest a link between our data and the possibility of observing and modelling urban geometric structures
using L\'{e}vy processes.
\end{abstract}
\end{frontmatter}

\section{Introduction}

The morphology of urban settlements and its dynamics has captured the interest
of physicists
\cite{Frankhauser94,Makse95,Benguigui95,Makse98,Schweitzer98,Malescio2000,Gomes2001}
as it may shed light on Zipf's law for cities
\cite{Manrubia97,Marsili98,Gabaix99,Malacarne01}, challenge theoretical
frameworks for cluster dynamics or improve predictions of urban growth
\cite{Makse95,Makse98,Schweitzer98}.

The search for a unified theory of urban morphology has focused on the premise
that cities can be conceptualized at several scales as fractals. At the
regional scale, rank-order plots of city size follow a fractal distribution
\cite{Frankhauser94} and population scales with city area as a power-law
\cite{Batty94}. More recently, it has been observed that the area distribution
of satellite cities, towns and villages around large urban centres also obeys
a power-law with exponent $\approx2$ \cite{Makse95,Makse98}. At the scale of
transportation networks, railway networks appear to have a fractal structure
\cite{Benguigui92}. At the scale of the neighbourhood, it has been suggested
that urban space resembles a Sierpinsky gasket \cite{Frankhauser94,Batty94}.
These scales are inter-related as summed up in \cite[pp. 241]{Frankhauser94}
(author's translation from French): '\emph{Polycentric growth, which is
connected to the non-homogeneous distribution of pre-urban cores and the birth
of a hierarchy of sub-centres, influences the morphology of the transport
network, which plays in itself an important role in axial growth and therefore
for the future spatial development of the urbanised area}'.

The fractal dimensions of US cities and international cities have values
ranging from $1.2778$ (Omaha, \cite{Shen2002}) to $1.93$ (Beijing,
\cite{Frankhauser94}), where the fractal dimension of large cities tends to
cluster around the latter value \cite{Frankhauser94,Batty94,Batty96,Shen2002}.
Studies of urban growth of London between $1820$ and $1962$ show that fractal
dimensions for this period vary from $1.322$ to $1.791$ \cite{Batty94}. The
fractal dimensions for the growth of Berlin in $1875$, $1920$ and $1945$ are
$1.43$, $1.54$ and $1.69$, respectively \cite{Frankhauser94}. The fractal
dimension of urban aggregates is a global measure of areal coverage, but
detailed measures of spatial distribution are clearly needed to complement
adequately the description of the morphology of an urban area
\cite{Schweitzer98}. Further, current approaches to data collection and
modelling identify cities as fractal only on the urban periphery of the giant
urban cluster that grows around the city core (or central business district),
as clusters become compact at distances close to the centre of the city
\cite{Batty94,Makse98}. Although remote sensing techniques are promising in
extracting urban morphology with greater detail, available studies have, to
our knowledge, been limited to individually selected, medium scale cities (see
e.g. \cite{Longley2000}).

Hillier and Hanson \cite{Hillier84} suggest an underlying structure to urban
open space that is determined by the complexity of buildings which bound the
space \cite{Batty1997}. Urban space available for pedestrian movement,
excluding by definition physical obstacles, is relatively linear. To extract
this linearity, a street can be approximated by a set of straight lines. The
global set of lines, the so-called \emph{axial map}, is defined as the least
number of longest straight lines. An axial map can be derived by drawing the
longest possible straight line on a city map, then the next longest line,
so-called axial line, until the open space is crossed and 'all axial lines
that can be linked to other axial lines without repetition are
linked'\ \cite{Hillier84,Jiang2002}. Figure \ref{axialmaps} shows several
axial maps.

Axial maps are relevant to urban theorists, as they encode the structure of
open space in urban settlements and provide a simplified signature of the
growth process. One could hope to extract the spatio-temporal dynamics of
axial map growth by analysing a sequence of aerial photographs of the urban
periphery over a period of growth. Conversely, one could hope to model urban
growth as the trajectory of $N$ walkers on a plane, the step of the walk being
axial line length.

Here we show that we can rescale axial line length and rank to obtain two
distinct rank-order curves that provide a classification for several cities
independently of city size. We also show that there is a class of cities that
does not obey this classification. The collapse of curves suggests that
spatial fluctuations in the length of urban linear structures, differing in
size and location, are governed by similar statistical rules and supports the
hypothesis that the linear dimension of large scale structures in cities
reflects generic properties of city growth \cite{Hillier2001}.

\section{Structure of urban space}

\subsection{Intermittency in urban space}

Let $\mathbf{l}_{i}=\left\{  \mathbf{l}_{i,j}\right\}  $, $j=1,\cdots,N_{i}$,
be the $N_{i}$ axial lines associated with city $i$. Each axial line,
$\mathbf{l}_{i,j}$ ($j=1,\cdots,N_{i}$) is defined by the coordinates of its
extremities
\[
\mathbf{l}_{i,j}=\left\{  \left(  x_{\left(  i,j\right)  ,1},y_{\left(
i,j\right)  ,1}\right)  ,\left(  x_{\left(  i,j\right)  ,2}\allowbreak
,y_{\left(  i,j\right)  ,2}\right)  \right\}
\]
The axial map of city $i$, $\mathcal{C}_{i}$, is thus a set of $N_{i}$ points
on a fourth dimensional space, $\mathcal{C}_{i}=\left\{  \left(  \rho
_{i,j},\theta_{i,j},l_{i,j},\varphi_{i,j}\right)  \right\}  $, where $\left(
\rho_{i},\theta_{i}\right)  $ are the polar coordinates of the axial line
geometric centre, $\mathbf{s}_{i,j}=\left(  \frac{x_{\left(  i,j\right)
,1}+x_{\left(  i,j\right)  ,2}}{2},\frac{y_{\left(  i,j\right)  ,1}+y_{\left(
i,j\right)  ,2}}{2}\right)  $, and $\left(  \pm\frac{l_{i,j}}{2},\varphi
_{i,j}\right)  $ are the polar coordinates of the axial line's extremities on
its geometric centre reference system, $\mathbf{d}_{i,j}\mathbf{=\pm}\left(
\frac{\left\vert x_{\left(  i,j\right)  ,1}-x_{\left(  i,j\right)
,2}\right\vert }{2},\frac{\left\vert y_{\left(  i,j\right)  ,1}-y_{\left(
i,j\right)  ,2}\right\vert }{2}\right)  $.

\begin{figure}
[ptb]
\begin{center}
\includegraphics[
height=6.2567in,
width=3.9932in
]%
{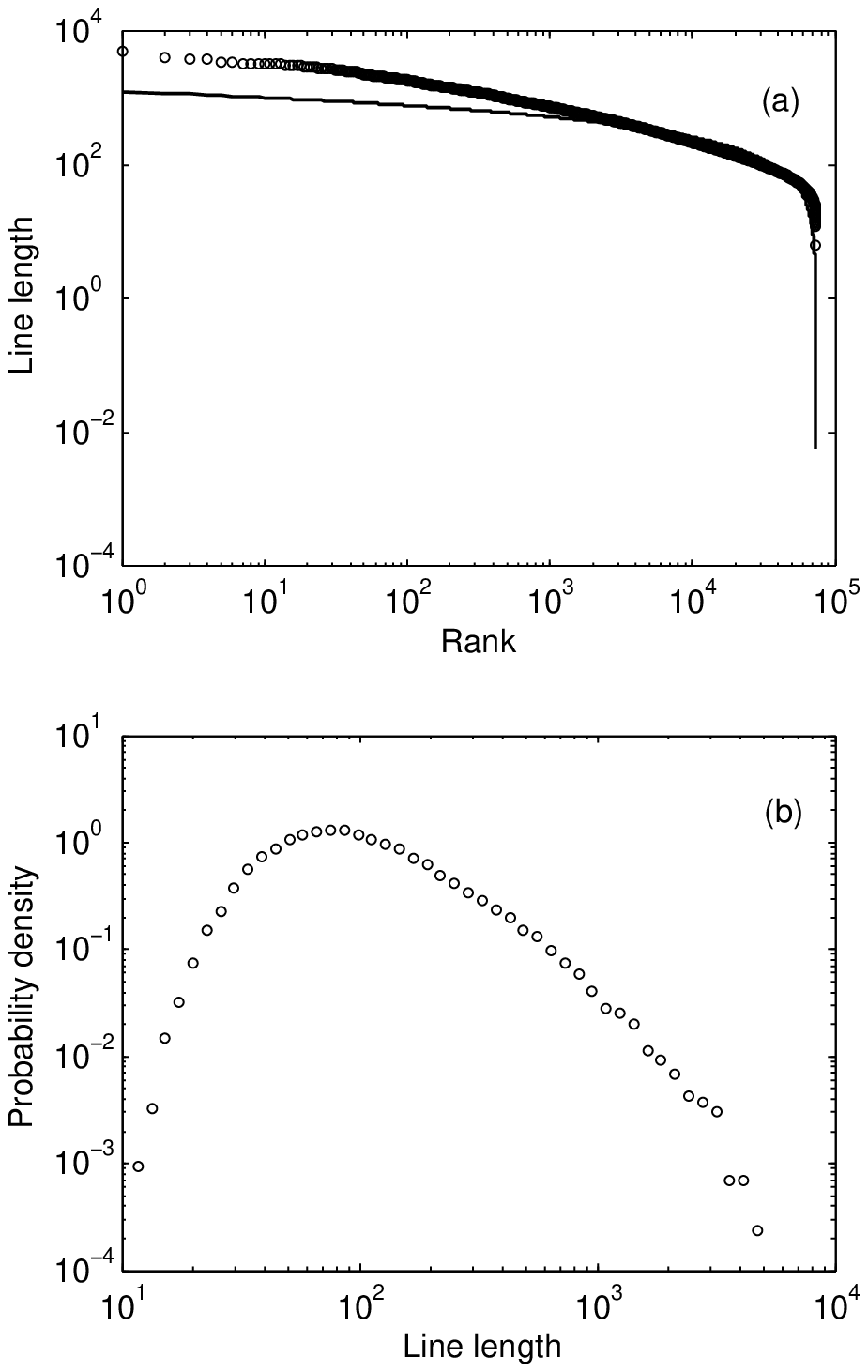}%
\caption{Data are shown for the city of Tokyo. (a) Rank-order plot of line
length (circle points) together with a fit of the data to a stretched
exponential pdf (solid line). (b) Unconditional probability density of line
length.}%
\label{tokyohist}%
\end{center}
\end{figure}

Coordinates $\rho$ and $\theta$ encode the geographic location of axial lines.
The unconditional distribution of $\varphi$ is multimodal for rather general
families of urban settlements. This occurs, for example, when land is
partitioned in clusters of randomly oriented orthogonal grids. Nevertheless,
the unconditional distribution of $l$ is unimodal and skewed to the right (see
Figure \ref{tokyohist}), and, thus, the only coordinate which is a good
candidate for inspection of intermittency in urban space. In what follows we
will analyse the statistics of line length for a database of cities. We start
by fitting the data for Tokyo to a stretched exponential distribution
\cite[pp. 153-154]{SornetteBook} in Figure \ref{tokyohist} \emph{(a)}, but
verify that the fit is unsuitable to describe the large events.

\subsection{Inverse square and cubic laws for the distribution of line length}

\begin{figure}
[ptb]
\begin{center}
\includegraphics[
height=3.7075in,
width=5.4241in
]%
{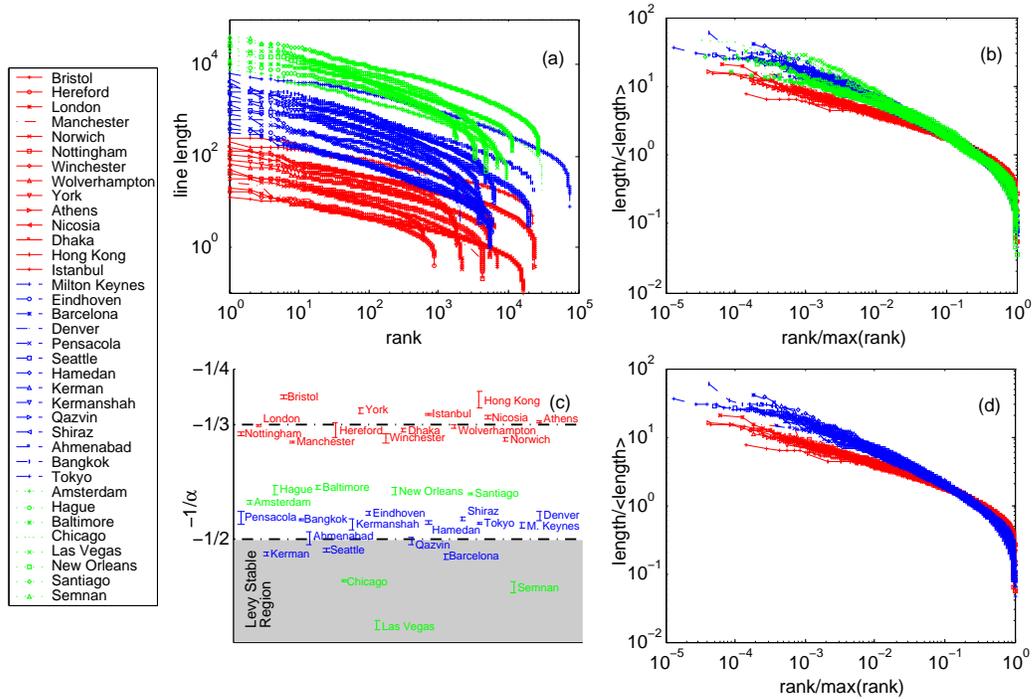}%
\caption{(a) Rank-order plot of line length versus rank. Consecutive curves
have been vertically shifted for clarity. (b) Data in (a) in scaled units. (c)
Exponents determined from least squares fits to the log-log data in (b) for
$10^{0.2}<y$. Error bars are $95\%$ confidence bounds. (d) Data in (b)
excluding cities in green. Cities are coloured according to their ordinate in
(c), and we have coloured a group of cities in green as they deviate
considerabily from the two universality classes in (d).}%
\label{Fig3}%
\end{center}
\end{figure}

\begin{figure}
[ptb]
\begin{center}
\includegraphics[
height=4.214in,
width=4.9056in
]%
{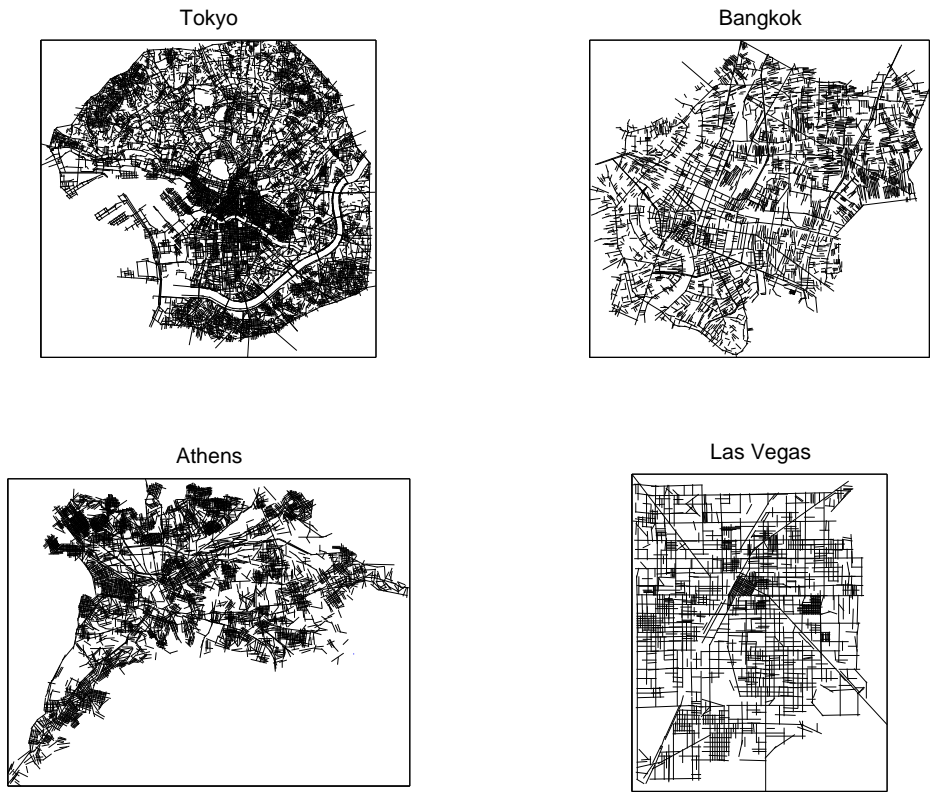}%
\caption{Axial maps of a sample from Figure \ref{Fig3} of cities coloured in
red (Athens), blue (Tokyo and Bangkok) and green (Las Vegas). Clockwise, the
probability of occurence of longer lines increases from bottom left (Athens,
local geometric structures), to upper graphs (Tokyo and Bangkok, global
geometric structures), to bottom right (Las Vegas, very strong global
geometric structures). }%
\label{axialmaps}%
\end{center}
\end{figure}

We analyse the unconditional probability distribution of (axial) line length
of $36$ cities in $14$ different countries (see Table \ref{Table1}). In our
analysis we use the rank-order technique \cite{SornetteBook}. To interpret the
apparently unsystematic data in Figure \ref{Fig3}(a) effectively, it is
instructive to scale the data. Since the rank ranges between $1$ and
$\max\left(  rank_{i}\right)  $, we define a scaled relative rank for city
$i$, $\widehat{r_{i,j}}\equiv rank_{j}/\max\left(  rank_{j}\right)  $ .
Similarly, for the ordinate, it is useful to define a scaled line length by
$\widehat{l_{i,j}}=l_{j}/\left\langle l_{j}\right\rangle $ \cite{Redner98}.

As shown in Figure \ref{Fig3}(d), there is relatively good collapse of the
data sets onto two master curves for $28$ of the $36$ cities under study (the
cities plot in red and blue). The other $8$ cities do not collapse clearly
onto a single curve (see Figure \ref{Fig3}(b)). Figure \ref{Fig3}(c) is a plot
of the exponents from a least-squares fit to the data of Figure \ref{Fig3}(b)
for $\log\widehat{l_{i,j}}>0.2$, where the data are visually the most linear.
The relatively small error bars ($95\%$ confidence bounds) in Figure
\ref{Fig3}(c) corroborate our choice of the cutoff. The fits on the rank order
plot lead to straight lines with slope $-1/\alpha$, which suggest that the
line length probability density may have a power-law tail, $P\left(  l\right)
\sim l^{-1-\alpha}$ with exponents close to $\alpha_{B}\simeq2$ (cities in
blue) or $\alpha_{R}\simeq3$ (cities in red). The inverse square and cubic
laws have diverging higher moments (larger than $2$ and $3,$ respectively) and
are not stable distributions.

Figure \ref{axialmaps} is a plot of several axial maps, where we only plot
lines with $\log\widehat{l_{i,j}}>0.2$ (the range of data used in the
least-squares fit of Figure \ref{Fig3}). We suggest that urban growth can be
regarded as a process where axial lines are added mainly to the urban
periphery (this could, in principle, be monitored through remote sensing
techniques for cities undergoing rapid urbanization) and modelled by $N$
walkers, which jump along the corresponding $N$ axial lines, extending the
city. More needs to be known on the distribution of the walkers' waiting time
to correctly model the dynamics of urban growth. Nevertheless, the walkers
would generate a non-stable process, as the exponents $\alpha_{B}$ and
$\alpha_{R}$ are, apparently, outside of the L\'{e}vy stable region.

\section{Discussion}

We have found that the length of urban open space structures displays
universal features, largely independent of city size, and is self-similar
across morphologically relevant \ ranges of scales ($2$ orders of magnitude)
with exponents $\alpha_{B}\simeq2$ (cities in blue) and $\alpha_{R}\simeq3$
(cities in red). Our results are unexpected as two universality classes appear
for a wide range of cities. The power-law tails of the pdfs support the
hypothesis that urban space has a fractal structure
\cite{Frankhauser94,Batty94}, but the parallelism to a Sierpinski gasket
\cite{Frankhauser94,Batty94} may be too simple for an accurate description.

Our findings show that it is important to model in detail the open space
geometry of urban aggregates. They also support the hypothesis that it is more
useful to model urban morphology as random rather than as the outcome of
rational decisions, as previously suggested \cite{Batty94,Makse95,Makse98}.

Cities with exponents $\alpha_{B}\simeq2$ (cities in blue) display open space
alignments which cross the whole structure, whilst cities with $\alpha
_{R}\simeq3$ (cities in red) tend not to. We propose that the large scale
linear structures of the two classes of cities can be explained by two
distinct non-stable L\'{e}vy processes, where the walker's jump has a tail
that goes as $\alpha_{B}\simeq2$ (cities in blue) or $\alpha_{R}\simeq3$
(cities in red). The walker's jumps are much larger for a process with
$\alpha_{B}\simeq2$ (cities in blue) than for a process with $\alpha_{R}%
\simeq3$ (cities in red), leading to the dominance of global geometric
structures for the former as opposed to local geometric structures for the
latter. The probability of a large axial line occurring for cities with
$\alpha_{G}<2$ (subset of cities in green) is larger than for the other cities
(see, e.g., Las Vegas in Figure \ref{Fig3}). We suggest that cities in green
with $2<\alpha_{G}<3$ have mixed influences of global and local structures and
that cities in green with $\alpha_{G}<2$ have very strong patterns of global
geometric structures. We note that the L\'{e}vy processes which, we propose,
imprint the dominant global geometric structures of a city may appear at a
late stage of urban growth, but change the city structure drastically (e.g. by
generating long axial transportation routes).

Physicists have found exponents $\alpha\approx3$ when studying the
distribution of normalized returns in financial markets, both for individual
companies \cite{Gopikrishnan98,Plerou99} and for market indexes
\cite{Gopikrishnan99}. Our results for the class of cities plot in red on
Figure \ref{Fig3} is reminiscent of these studies. We believe that parallels
between urban growth and finance may not be too far fetched, as both processes
seem to be largely dominated by geometric phenomena
\cite{Malescio2000,Stanley01}. Indeed, there may be similarities between the
dynamics of price fluctuations and urban growth, and we propose that axial
lines may be seen as the urban equivalent of economic returns.

As more data becomes available through remote sensing, quantitative analyses
should provide an improved view of the spatio-temporal dynamics of urban
growth, particularly in squatter settlements, where time-scales for growth are
much shorter than in conventional cities and one could hope to model growth
against observed data. Further studies are still required, but it seems that
the impact of local controls on growth (e.g. the green belt policy for London)
is, at most, spatially localized. Indeed, at a 'macro' level, cities display a
surprising degree of universality.

\section{Acknowledgments}

We acknowledge financial support from Grant EPSRC GR/N21376/01. We thank Prof.
Bill Hillier for most useful discussions and encouragement, Prof. Mike Batty
for suggestions on the rank-order plots and Shinichi Iida for invaluable
insight on the axial map of Tokyo.

\bibliographystyle{unsrt}
\bibliography{acompat,Papers}

\newpage%

\begin{table}[tbp] \centering
\caption{\label{Table1}Geographical location and number of lines of the
cities analysed.}%

\begin{tabular}
[c]{lll}\hline
\textbf{Country} & \textbf{City} & \textbf{Number of lines}\\\hline
Japan & Tokyo & $73753$\\[-2ex]%
U.S.A. & Chicago & $30571$\\[-2ex]%
Chile & Santiago & $26821$\\[-2ex]%
Thailand & Bangkok & $24223$\\[-2ex]%
Greece & Athens & $23329$\\[-2ex]%
Turkey & Istanbul & $21798$\\[-2ex]%
U.S.A. & Seattle & $20213$\\[-2ex]%
U.K. & London & $15969$\\[-2ex]%
U.S.A. & Baltimore & $11636$\\[-2ex]%
Netherlands & Amsterdam & $9619$\\[-2ex]%
U.K. & Bristol & $7028$\\[-2ex]%
U.S.A. & Las Vegas & $6909$\\[-2ex]%
Iran & Shiraz & $6258$\\[-2ex]%
Cyprus & Nicosia & $6023$\\[-2ex]%
Netherlands & Eindhoven & $5782$\\[-2ex]%
U.K. & Milton Keynes & $5581$\\[-2ex]%
Spain & Barcelona & $5575$\\[-2ex]%
U.K. & Wolverhampton & $5423$\\[-2ex]%
India & Ahmenabad & $4876$\\[-2ex]%
U.S.A. & New Orleans & $4846$\\[-2ex]%
Iran & Kerman & $4372$\\[-2ex]%
U.K. & Nottingham & $4365$\\[-2ex]%
U.K. & Manchester & $4308$\\[-2ex]%
U.S.A. & Pensacola & $4296$\\[-2ex]%
Iran & Hamadan & $3855$\\[-2ex]%
Iran & Qazvin & $3723$\\[-2ex]%
Netherlands & The Hague & $3350$\\[-2ex]%
U.K. & Norwich & $2119$\\[-2ex]%
U.S.A. & Denver & $2092$\\[-2ex]%
Iran & Kermanshah & $1870$\\[-2ex]%
U.K. & York & $1773$\\[-2ex]%
Iran & Semnan & $1770$\\[-2ex]%
Bangladesh & Dhaka & $1566$\\[-2ex]%
Hong Kong & Hong Kong & $916$\\[-2ex]%
U.K. & Hereford & $854$\\[-2ex]%
U.K. & Winchester & $616$\\\hline
\end{tabular}
%

\end{table}%
%

%

\end{document}